\documentclass[runningheads]{llncs}
\usepackage{amsmath}
\usepackage{amsfonts}
\usepackage{dsfont}
\usepackage{graphicx}
\usepackage{tikz}
\usepackage{multirow}
\usepackage{array}
\usepackage{tabularx,booktabs}
\usepackage{subcaption}
\usepackage{dirtytalk}
\usepackage{bm}
\usepackage{dsfont}
\usepackage{float}
\usetikzlibrary{arrows.meta}
\usepackage{color}
\usepackage{hyperref}

\begin{document}


\title{Exploring Patient Data Requirements in Training Effective AI Models for\\MRI-based Breast Cancer Classification}

\titlerunning{Exploring Patient Data Requirements for 
Breast Cancer Classification}

\author{Solha Kang\inst{1} \and
Wesley De Neve\inst{1,3} \and
Francois Rameau\inst{2} \and
Utku Ozbulak\inst{1,3}}

\authorrunning{Kang et al.}

\institute{
Center for Biosystems and Biotech Data Science,\\ Ghent University Global Campus, Republic of Korea
\and
The State University of New York Korea, Republic of Korea
\and
Department of Electronics and Information Systems,\\ Ghent University, Belgium\\
\email{utku.ozbulak@ghent.ac.kr}
}

\maketitle

\begin{abstract}
The past decade has witnessed a substantial increase in the number of startups and companies offering AI-based solutions for clinical decision support in medical institutions. However, the critical nature of medical decision-making raises several concerns about relying on external software. Key issues include potential variations in image modalities and the medical devices used to obtain these images, potential legal issues, and adversarial attacks. Fortunately, the open-source nature of machine learning research has made foundation models publicly available and straightforward to use for medical applications. This accessibility allows medical institutions to train their own AI-based models, thereby mitigating the aforementioned concerns. Given this context, an important question arises: how much data do medical institutions need to train effective AI models? In this study, we explore this question in relation to breast cancer detection, a particularly contested area due to the prevalence of this disease, which affects approximately 1 in every 8 women. Through large-scale experiments on various patient sizes in the training set, we show that medical institutions do not need a decade's worth of MRI images to train an AI model that performs competitively with the state-of-the-art, provided the model leverages foundation models. Furthermore, we observe that for patient counts greater than 50, the number of patients in the training set has a negligible impact on the performance of models and that simple ensembles further improve the results without additional complexity.
\end{abstract}

\section{Introduction}

Although AI-based models that utilize deep neural networks have demonstrated successful results in various medical imaging tasks, such as tumor detection~\cite{derma_skin_cancer_detection}, organ segmentation~\cite{organ_segmentation}, and disease classification~\cite{disease_Cls1,disease_Cls2}, these models have not yet been widely adopted in clinical settings due to several factors that limit their usage~\cite{clinical_challenge}. First, the critical nature of medical decision-making demands exceptionally high accuracy and reliability, as errors can have severe consequences for patient health. Consequently, developing AI algorithms for medical applications requires not only a deep understanding of both computer science and medicine but also compliance with various regulations and standards, such as the Health Insurance Portability and Accountability Act (HIPAA)~\cite{act1996health}. This complexity makes it challenging for private companies to develop and commercialize their products effectively.

Second, the lack of standardized and high-quality medical data complicates the training and validation of AI models. Variations in imaging modalities, the use of different medical devices, and inconsistencies in data annotation all contribute to this complexity~\cite{liu2022medical}. For example, although breast cancer is one of the most researched types of cancer, the majority of studies investigating breast cancer via MRI or mammography imaging use private datasets, making comparisons across various methods nearly impossible~\cite{brca_mri}.

\begin{figure}[t!]
    \centering
    \includegraphics[width=0.75\textwidth]{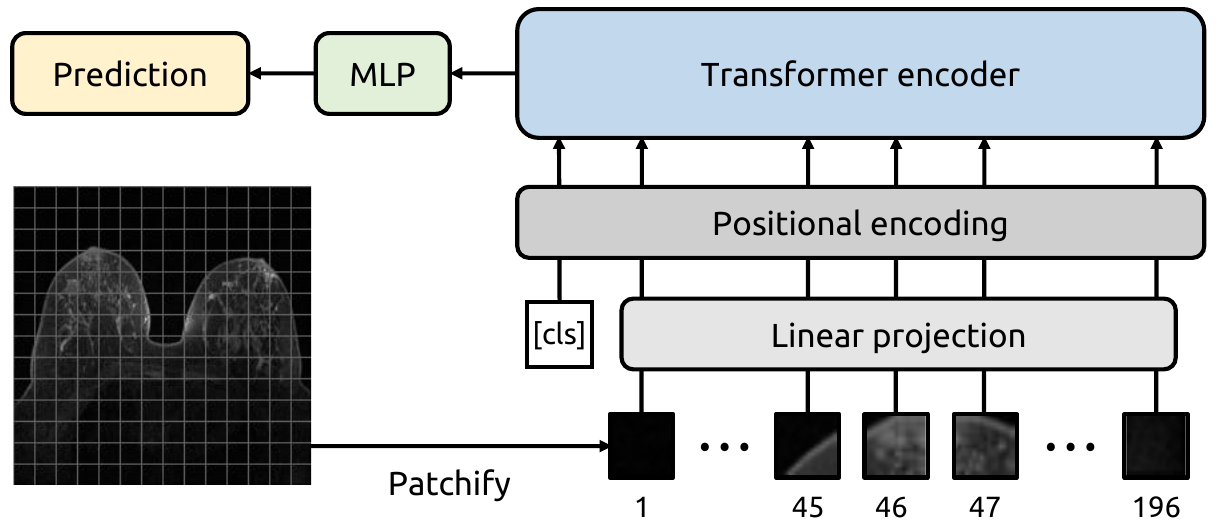}
    \caption{Visualization of the ViT architecture and image patch tokenization.}
    \label{fig:transformer_fig}
\end{figure}

Lastly, concerns about data privacy and security receive a lot of attention in the medical field. Various adversarial attacks, such as data poisoning attacks, which can taint training datasets, pose substantial risks~\cite{medical_poisoning,medical_adv_attack}. The potential for such attacks introduces legal and ethical issues that further hinder the deployment of AI in healthcare settings.

All of the aforementioned factors contribute to the decision of medical institutions, such as university hospitals, to develop their own AI-based solutions, typically in collaboration with computer science departments~\cite{maleki2024role,mekki2024physicians}. Until recently, such efforts required tremendous amounts of data and expertise, as the models had to be specifically created for each task at hand. This posed considerable challenges in terms of resources and specialized knowledge, often limiting the ability of smaller institutions to develop effective AI solutions~\cite{iliashenko2019opportunities}.

The emergence of foundation models~\cite{bommasani2021opportunities}, which are mostly based on the transformer architecture, has been a game changer for medical AI, including breast cancer detection~\cite{awais2023foundational}. These models, pre-trained on vast amounts of diverse data, provide a robust and versatile base upon which researchers can build their applications~\cite{dino}. For a comprehensive overview on foundation models, we refer the interested readers to the work of~\cite{bommasani2021opportunities}. By leveraging these foundation models in the form of transfer learning, medical institutes can substantially reduce the amount of data and time needed to develop accurate and reliable AI solutions. This not only democratizes access to advanced AI technologies but also enhances the feasibility of creating in-house solutions tailored to specific medical needs~\cite{azad2023foundational}.

Given the ability of such pretrained models to discover relationships with potentially fewer images, we ask the following question: how many patients would a medical institute need to create a model comparable to the state-of-the-art if they utilize foundation models? Specifically, can a medical institute avoid the need to spend a decade collecting and annotating MRI data to train an effective breast cancer detection model? To answer this question, we employ the DUKE Breast Cancer Dataset~\cite{duke_dataset}, curated over 14 years, aimed at detecting malignant breast tumors in MRI data. Through large-scale experiments, we investigate and identify the diminishing returns of image quantities in the training set for breast cancer detection via MRI data, and we make the following contributions:

\begin{itemize}
    \item We demonstrate that pretrained models can achieve state-of-the-art results in MRI-based breast cancer detection even with a limited number of images, showcasing their efficiency in data-scarce scenarios.
    \item We observe minimal performance differences among various pretraining methods, indicating that the choice of pretraining technique may have a negligible impact on the overall outcomes.
    \item We observe that the selection of patients in the training set has a negligible impact on the performance of models for patient counts as few as 50.
    \item We experimented with simple ensemble methods and demonstrated their effectiveness in improving overall performance without adding substantial complexity.
\end{itemize}

\begin{figure}[t!]
\centering
\begin{subfigure}{0.48\textwidth}
\centering
\includegraphics[width=0.45\textwidth]{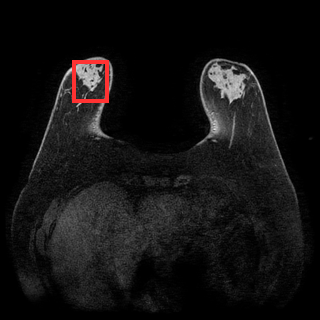}
\includegraphics[width=0.45\textwidth]{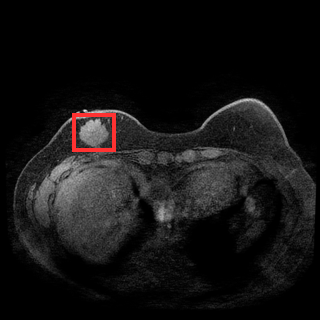}
\includegraphics[width=0.45\textwidth]{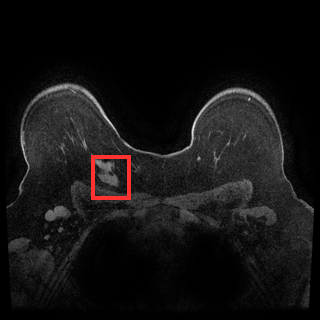}
\includegraphics[width=0.45\textwidth]{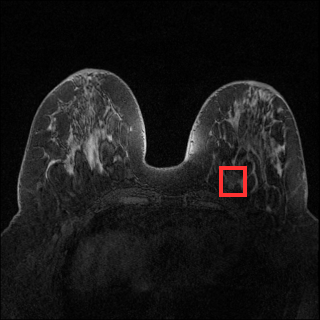}
\includegraphics[width=0.45\textwidth]{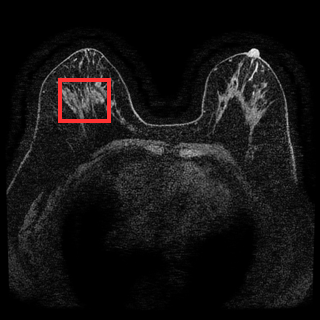}
\includegraphics[width=0.45\textwidth]{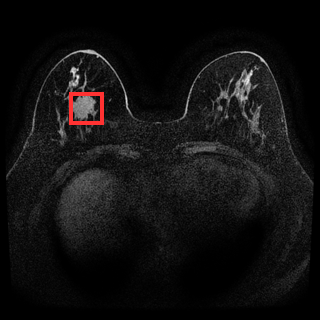}
\caption{Tumor-positive}
\label{fig:duke_pos}
\end{subfigure}
\begin{subfigure}{0.48\textwidth}
\centering
\includegraphics[width=0.45\textwidth]{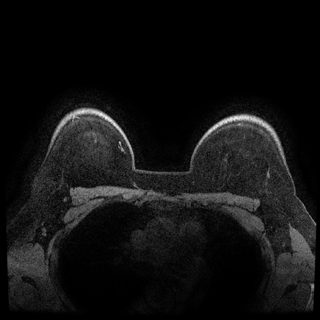}
\includegraphics[width=0.45\textwidth]{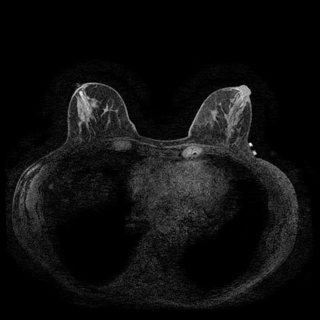}
\includegraphics[width=0.45\textwidth]{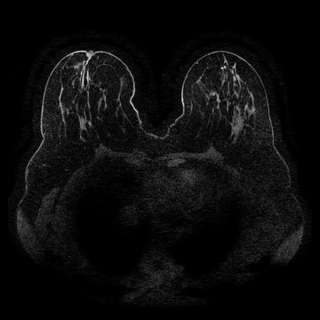}
\includegraphics[width=0.45\textwidth]{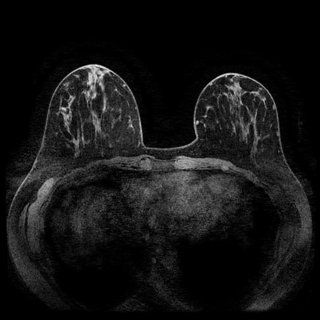}
\includegraphics[width=0.45\textwidth]{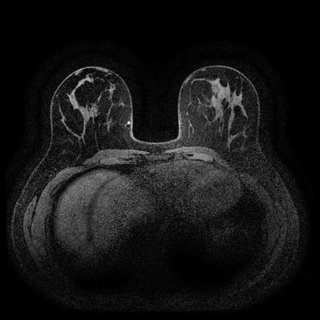}
\includegraphics[width=0.45\textwidth]{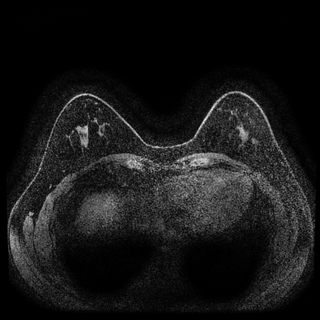}
\caption{Tumor-negative}
\label{fig:duke_neg}
\end{subfigure}
\caption{An example set of images from the Duke breast MRI dataset: (a) Tumor-positive breast MRI images, overlaid with bounding boxes indicating tumor locations, and (b) Tumor-negative breast MRI images.}
\label{fig:duke_examples}
\end{figure}

\section{Methodology}

In this section, we provide a description of the dataset used, outline the models utilized, and describe the proposed method in detail.

\subsection{Models}
\label{sec:models}

To investigate the impact of data scarcity on model training, we employ the most commonly used transformer-based computer vision architecture, Vision Transformer-Base/16 (ViT-B/16)~\cite{vit}. Vision transformers are known to be data-hungry and perform poorly compared to convolutional neural networks in data-scarce regimes. To overcome this limitation, numerous self-supervised learning (SSL) frameworks have been proposed in the past couple of years~\cite{mocov1,dino,mae}. For a detailed overview of SSL frameworks, we refer interested readers to the following surveys~\cite{khan2022contrastive_survey,ozbulak2023know}. Among these SSL frameworks, we employ two models pretrained using (1) DINO~\cite{dino}, a discriminative SSL framework, and (2) MAE~\cite{mae}, which relies on a generative approach. Pretraining for these models is performed on the ImageNet dataset~\cite{ILSVRC15:rus}, a large-scale dataset containing natural images. In addition to these two models, we also experiment with two additional ViTs: one that is randomly initialized and another that is pretrained in a supervised fashion on the same (ImageNet) dataset. Models that are pretrained using the aforementioned methods on large datasets typically serve as foundation models. Note that fine-tuning a foundation model for a specific dataset falls under the category of transfer learning and has been studied extensively for the past decade~\cite{transfer_learning_1}.

As suggested by~\cite{vit}, we use ViTs for images of size $224 \times 224$ with image tokens/patches of size $16 \times 16$, resulting in a total of $196$ tokens. We modify the final linear layer of the model to accommodate the two-class classification problem tackled in this work.

\subsection{Dataset}
\label{sec:dataset}

The Duke Breast Cancer Dataset, sourced from 922 patients with invasive breast cancer treated at Duke Hospital over a span of 14 years, stands as one of the largest publicly available datasets of 3D MRI breast cancer images~\cite{duke_dataset}. The primary task with this dataset is to determine the presence of breast tumors in the MRI images based on 2D slices. For 3D to 2D conversion, we follow the procedure detailed in~\cite{duke_intrinsic,duke_dataset_property}. When the 3D images of this dataset are separated into 2D slices, this operation results in more than $100,000$ 2D images corresponding to approximately $130$ images for each patient. Due to the large quantity of data, the majority of research efforts employing this dataset use only a subset of these images to showcase experimental results. Unlike those efforts, we use the entire DUKE dataset for a comprehensive analysis and incorporate a data-splitting methodology to investigate potential disparities in patient selection for the training set. Example images from the DUKE dataset are displayed in \figurename~\ref{fig:duke_examples}, highlighting regions with tumors in tumor-positive images, thereby showcasing the complexity of the tackled problem.

\begin{figure}[t!]
    \centering
    \includegraphics[width=0.98\textwidth]{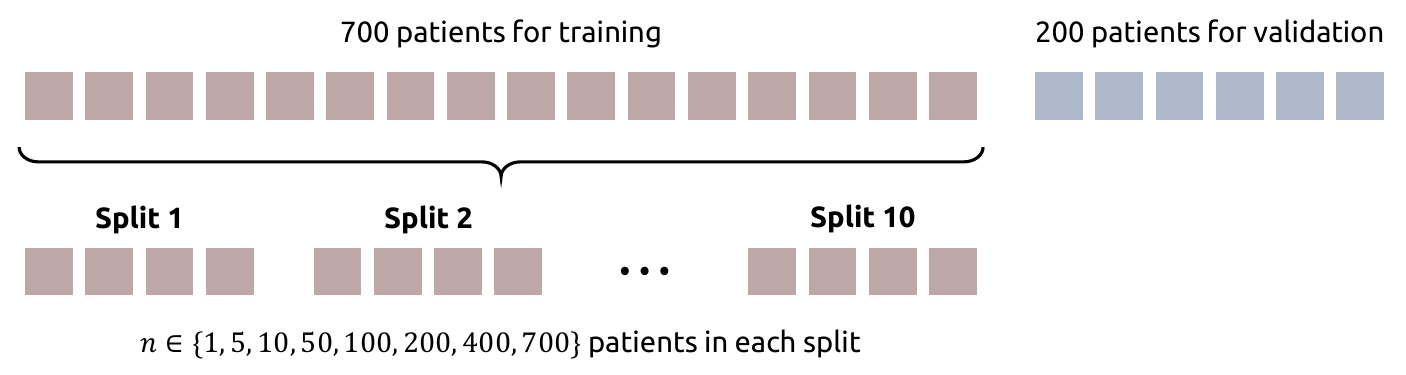}
    \caption{Overview of the generation of training splits. A fixed validation set of 200 patients is randomly sampled from a total of 900 patients. From the remaining 700 patients, 10 training splits of $n \in \{1, 5, 10, 50, 100, 200, 400, 700\}$ patients are randomly sampled for each patient count.}
    \label{fig:tr_split}
\end{figure}

\textbf{Patient-based data splitting for consistent evaluation}\,\textendash\,To ensure consistent evaluation, we first sample a fixed validation set consisting of images obtained from 200 patients. Then, to perform a thorough investigation that accounts for potential disparities in the selected patients in the training set, we employ stratified sampling to create multiple training subsets with varying numbers of patients. Specifically, we randomly sample images from 1, 5, 10, 50, 100, 200, 400, and 700 patients from the cohort of the remaining 722 patients and generate ten unique training splits per patient count. This results in a total of 80 unique training datasets (8 different patient counts $\times$ 10 unique splits). For each patient count, the ten training splits contain an equal number of positive and negative images. A visual description of the patient splitting technique we employ can be found in \figurename~\ref{fig:tr_split}.

The aforementioned approach provides a comprehensive set of training datasets for model evaluation and enables us to assess potential discrepancies in model performance with respect to the selection of patients within the training set.

\subsection{Training}
\label{sec:training}

We train our foundation models using a grid-search strategy, applying the SGD optimizer with learning rates of 0.1, 0.01, and 0.001, paired with weight decay settings of either 0 or 0.0001. Our training process employs the Cross-Entropy Loss function, with a batch size of 32. We adopt a cosine annealing learning rate scheduler in alignment with the research on downstream transferability of self-supervised models. We experiment with a momentum of 0 and 0.1. For augmentation, we exclusively use random resized crops of size $224 \times 224$ to maintain the fidelity of MRI images. Additionally, early stopping is implemented in all training runs, with a patience threshold set at 5 epochs. Based on the aforementioned grid-search approach, we select models with the highest validation performances on their respective data splits to demonstrate experimental results.

\begin{figure}[t!]
\centering
\includegraphics[width=0.99\textwidth]{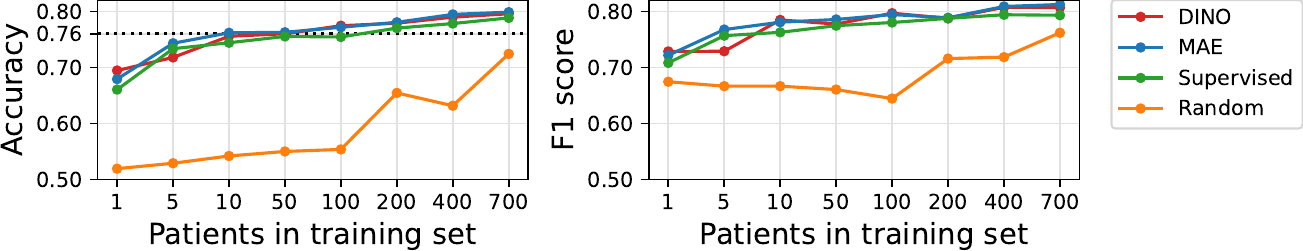}
\caption{(left) Validation accuracies of the best-performing ViT models trained with 1, 5, 10, 50, 100, 200, 400, 700 patients across all 10 training splits, and (right) corresponding F1-scores for the selected models.}
\label{fig:exp_results_line}
\end{figure}

\section{Experimental Results}

\begin{figure}[t!]
\centering
\begin{subfigure}{\textwidth}
\begin{subfigure}{0.5\textwidth}
\includegraphics[width=0.99\textwidth]{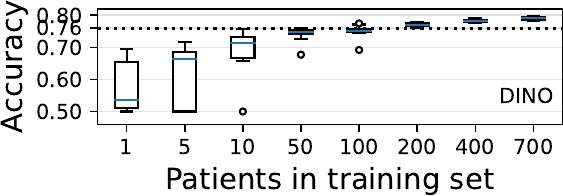}
\end{subfigure}
\begin{subfigure}{0.5\textwidth}
\includegraphics[width=0.99\textwidth]{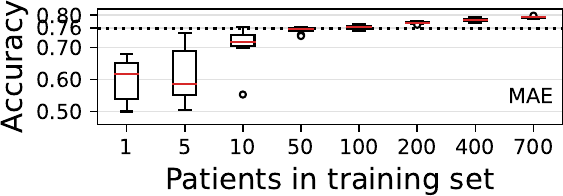}
\end{subfigure}
\begin{subfigure}{0.5\textwidth}
\includegraphics[width=0.99\textwidth]{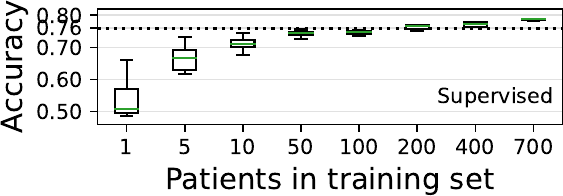}
\end{subfigure}
\begin{subfigure}{0.5\textwidth}
\includegraphics[width=0.99\textwidth]{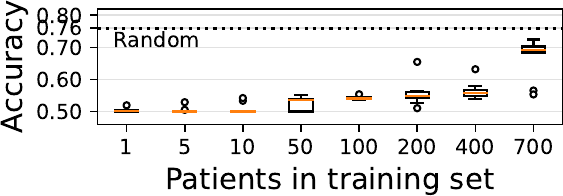}
\end{subfigure}
\caption{Accuracy on the validation set}
\end{subfigure}
\begin{subfigure}{\textwidth}
\begin{subfigure}{0.5\textwidth}
\includegraphics[width=0.99\textwidth]{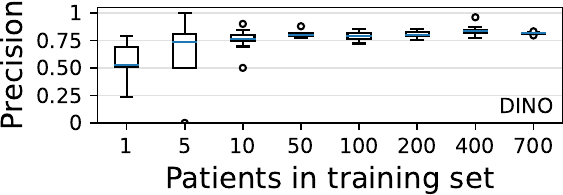}
\end{subfigure}
\begin{subfigure}{0.5\textwidth}
\includegraphics[width=0.99\textwidth]{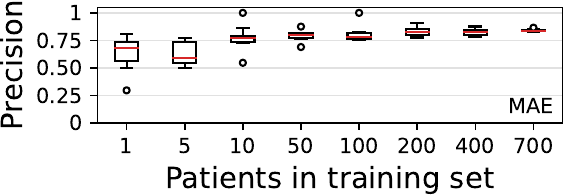}
\end{subfigure}
\begin{subfigure}{0.5\textwidth}
\includegraphics[width=0.99\textwidth]{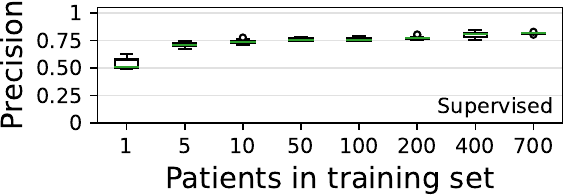}
\end{subfigure}
\begin{subfigure}{0.5\textwidth}
\includegraphics[width=0.99\textwidth]{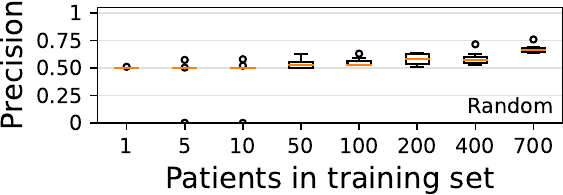}
\end{subfigure}
\caption{Precision on the validation set}
\end{subfigure}
\caption{Box plots illustrating the distribution of (a) accuracy as well as (b) precision on validation sets across ViT models trained with 1, 5, 10, 50, 100, 200, 400, and 700 patients. For each patient count in the training set, patients are randomly sampled from the training dataset to create 10 training splits.}
\label{fig:exp_results_boxplot}
\end{figure}

As detailed in Section~\ref{sec:models}, we train four ViT-B/16 models where three of those models: three pretrained on the ImageNet dataset using DINO, MAE, and supervised training, and a fourth model with random initialization. We train those models with the grid-search approach detailed in Section~\ref{sec:training} and present results in \figurename~\ref{fig:exp_results_line} and \figurename~\ref{fig:exp_results_boxplot}.

Specifically, \figurename~\ref{fig:exp_results_line} displays the effectiveness of the best-performing models on the validation set, measured by accuracy and F1-score, for each patient split. To quantify the influence of selecting different patients for training, we select best-performing models on their respective patient splits and represent the distribution of accuracies in \figurename~\ref{fig:exp_results_boxplot}. In figures comparing the accuracy, the dashed line represents the state-of-the-art results obtained from the work of~\cite{duke_intrinsic}. Based on these results, we make the observations listed below.

\textbf{Foundation models do not require substantial MRI training data for state-of-the-art results}. Unsurprisingly, we observe a steady improvement in model performance with an increased number of patients in the training set. Moreover, we find that in the best-case scenario, having a training set consisting of only 10 patients and using foundation models is enough to reach the accuracy obtained in the work of~\cite{duke_intrinsic}. From patient counts of 50 to 700, a 14-fold increase in the patient count only leads to approximately a $4\%$ increase in accuracy, suggesting that there is a point at which increasing the number of patients in the training set yields diminishing returns in performance enhancement.

\textbf{The pre-training method used has minimal impact on model performance}.
As shown in \figurename~\ref{fig:exp_results_line} and \figurename~\ref{fig:exp_results_boxplot}, the models exhibit a similar increase in performance regardless of the training method. This suggests that the specific pre-training method used for the models might not significantly impact their performance. This finding highlights the robustness of the pre-trained models, indicating that once a certain amount of training data is reached, the choice of pre-training approach becomes less critical to the model's success.

\textbf{Patient selection for training data is not critical}.
In \figurename~\ref{fig:exp_results_boxplot}, we illustrate the distribution of the highest validation accuracies as well as precision scores obtained for each training split. We observe that, especially for the three foundation models, the variability across different splits decreases with the increase in the number of patients used for training. Contrary to the findings of~\cite{med_data_critical}, for patient counts of 50 and onward, all training splits achieve similar performances, indicating that patients in the training set do not have a noticeable impact on model performance.

\textbf{Simple ensembles increase the performance}.
Ensemble models in machine learning are generally known to have improved performance over the individual models that comprise the ensembles~\cite{hansen1990neural}. Utilizing the three best-performing pre-trained models for individual splits, we experimented with a simple ensemble using majority voting for the case of 50 patients in the training data, provided accuracy values and F1-scores in Table~\ref{tbl:ensemble}. As can be seen, for 9 out of 10 splits, the ensemble model achieves higher accuracy compared to the best-performing individual model, highlighting the potential for improved diagnostic accuracy in medical applications with minimal additional complexity.

\begin{table}[t]
\centering
\scriptsize
\caption{For 10 splits containing only the data of 50 patients in the training set, the performance of three foundation models and a simple majority vote-based ensemble model of the three foundation models are provided. For each row, the highest accuracy value and F1-score are highlighted with \underline{underlined text} and \textbf{bold text}, respectively.}
\label{tbl:ensemble}
\begin{tabular}{ccccccccc}
\cmidrule[0.75pt]{2-9}
~ & \multicolumn{2}{c}{DINO} & \multicolumn{2}{c}{MAE} & \multicolumn{2}{c}{Supervised} & \multicolumn{2}{c}{Ensemble} \\
\cmidrule[0.75pt]{1-9}
Data Split & \phantom{-}Accuracy\phantom{-} & \phantom{-}F1\phantom{-} & \phantom{-}Accuracy\phantom{-} & \phantom{-}F1\phantom{-} &\phantom{-}Accuracy\phantom{-} & \phantom{-}F1\phantom{-} &\phantom{-}Accuracy\phantom{-} & \phantom{-}F1\phantom{-} \\
\midrule
1 & 0.760 & 0.776 & 0.759 & 0.789 & 0.749 & 0.775 & \underline{0.765} & \textbf{0.791} \\
2 & 0.743 & 0.763 & 0.720 & 0.763 & 0.729 & 0.750 & \underline{0.756} & \textbf{0.783} \\
3 & 0.757 & 0.778 & 0.763 & 0.779 & 0.744 & 0.765 & \underline{0.766} & \textbf{0.787} \\
4 & 0.755 & 0.788 & 0.761 & 0.789 & 0.752 & 0.772 & \underline{0.768} & \textbf{0.797} \\
5 & 0.740 & 0.766 & \underline{0.753} & 0.775 & 0.725 & 0.758 & 0.749 & \textbf{0.776} \\
6 & 0.763 & 0.774 & 0.770 & \textbf{0.791} & 0.746 & 0.754 & \underline{0.770} & 0.785 \\
7 & 0.751 & 0.760 & 0.751 & 0.769 & 0.733 & 0.767 & \underline{0.760} & \textbf{0.784} \\
8 & 0.756 & 0.770 & 0.766 & \textbf{0.790} & 0.741 & 0.759 & \underline{0.766} & 0.785 \\
9 & 0.735 & 0.756 & 0.759 & 0.769 & 0.713 & 0.708 & \underline{0.759} & \textbf{0.771} \\
10 & 0.750 & 0.763 & 0.763 & \textbf{0.791} & 0.755 & 0.766 & \underline{0.769} & 0.789 \\
\bottomrule
\end{tabular}
\end{table}

\section{Conclusions}

In this research effort, we demonstrated that medical institutes that perform MRI-based breast cancer detection do not need a decade of data curation to train an effective AI model that performs comparably to the state-of-the-art. In particular, we showed that minimal training data, requiring as few as 50 patients, can match previous accuracy benchmarks, with ensemble methods further improving the results.

Our goal with this research effort is to encourage medical institutes to adopt AI-based diagnostic tools more rapidly by demonstrating the feasibility of achieving high accuracy with limited amounts of data. Indeed, approaches that take advantage of foundation models can substantially reduce the time and resources required for data collection, enabling faster implementation of advanced diagnostic technologies and improving patient outcomes. We hope that our findings will drive innovation in medical diagnostics and contribute to more widespread and equitable access to state-of-the-art healthcare solutions.

That being said, we would like to highlight that observations made in this work may be partial to the employed dataset. As such, we believe that a future work exploring a wider-range of datasets is necessary to solidify our findings.

Although we demonstrated the diminishing returns of image quantities on the models' performance, it is nevertheless more beneficial to employ all existing data to make marginal gains in model performance. As such, we believe future research efforts investigating the optimal amount of data, training time, and performance would greatly benefit the medical AI community.

\bibliographystyle{splncs04}
\bibliography{main}
\end{document}